\documentclass[11pt]{article}
\usepackage{amsmath,amssymb,cite}

\makeatletter
\@addtoreset{equation}{section}
\makeatother

\def\crampest{\medmuskip = 1mu plus 1mu minus 1mu}

\def\ben{\begin{equation}}
\def\een{\end{equation}}

 \let\m=\mu \let\n=\nu  \let\p=\pi

\let\C=\Chi

\def\nn{\nonumber} \def\bd{\begin{document}} \def\ed{\end{document}}
\def\ds{\documentstyle} \let\fr=\frac \let\bl=\bigl \let\br=\bigr
\let\Br=\Bigr \let\Bl=\Bigl
\let\bm=\bibitem
\let\na=\nabla
\let\pa=\partial \let\ov=\overline
\newcommand{\be}{\begin{equation}}
\newcommand{\ee}{\end{equation}}
\def\ba{\begin{array}}
\def\ea{\end{array}}
\def\ft#1#2{{\textstyle{\frac{\scriptstyle #1}{\scriptstyle #2} } }}
\def\fft#1#2{{\frac{#1}{#2}}}
\def\del{\partial}
\def\vp{\varphi}
\def\sst#1{{\scriptscriptstyle #1}}
\def\oneone{\rlap 1\mkern4mu{\rm l}}
\def\td{\tilde}
\def\wtd{\widetilde}
\def\ie{{\it i.e.\ }}
\def\dalemb#1#2{{\vbox{\hrule height .#2pt
        \hbox{\vrule width.#2pt height#1pt \kern#1pt
                \vrule width.#2pt}
        \hrule height.#2pt}}}
\def\square{\mathord{\dalemb{6.8}{7}\hbox{\hskip1pt}}}
\newcommand{\ho}[1]{$\, ^{#1}$}
\newcommand{\hoch}[1]{$\, ^{#1}$}
\newcommand{\bea}{\begin{eqnarray}}
\newcommand{\eea}{\end{eqnarray}}
\newcommand{\ra}{\rightarrow}
\newcommand{\lra}{\longrightarrow}
\newcommand{\Lra}{\Leftrightarrow}
\newcommand{\bp}{\tilde \beta^\prime}
\newcommand{\tr}{{\rm tr} }
\newcommand{\Tr}{{\rm Tr} }
\def\0{{\sst{(0)}}}
\def\1{{\sst{(1)}}}
\def\2{{\sst{(2)}}}
\def\3{{\sst{(3)}}}
\def\4{{\sst{(4)}}}
\def\5{{\sst{(5)}}}
\def\6{{\sst{(6)}}}
\def\7{{\sst{(7)}}}
\def\8{{\sst{(8)}}}
\def\m{{\sst{(m)}}}
\def\n{{\sst{(n)}}}
\def\cA{{{\cal A}}}
\def\cB{{{\cal B}}}
\def\cF{{{\cal F}}}
\def\cG{{{\cal G}}}
\def\cH{{{\cal H}}}
\def\tV{\widetilde V}
\def\tW{\widetilde W}
\def\tH{\widetilde H}
\def\tE{\widetilde E}
\def\tF{\widetilde F}
\def\tA{\widetilde A}
\def\im{{{\rm i}}}
\def\tY{{{\wtd Y}}}
\def\ep{{\epsilon}}
\def\vep{{\varepsilon}}
\def\bD{{{\bar D}}}
\def\R{{{\mathbb R}}}
\def\C{{{\mathbb C}}}
\def\H{{{\mathbb H}}}
\def\CP{{{\mathbb C}{\mathbb P}}}
\def\RP{{{\mathbb R}{\mathbb P}}}
\def\Z{{{\mathbb Z}}}
\def\bA{{{\mathbb A}}}
\def\bB{{{\mathbb B}}}
\def\bC{{{\mathbb C}}}
\def\bD{{{\mathbb D}}}
\def\bE{{{\mathbb E}}}
\def\bZ{{{\mathbb Z}}}
\def\Re{{{\frak{Re}}}}
\def\Im{{{\frak{Im}}}}
\def\cosec{{\,\hbox{cosec}\,}}
\def\Gm{{\Gamma_{\!\! -}}}
\def\Gp{{\Gamma_{\!\! +}}}
\def\stan{{standard }}
\def\nonstan{{supernumerary }}
\def\p{{\partial}}
\def\kdel#1{{\fft{\del}{\del#1}}}

\def\bog{{Bogomolny }}
\def\om{{\omega}}

\newcommand{\tamphys}{\it George and Cynthia Woods Mitchell  Institute
for Fundamental Physics and Astronomy,\\
Texas A\&M University, College Station, TX 77843, USA}

\newcommand{\auth}{
H. L\"u\hoch{\dagger},
Jianwei Mei\hoch{\dagger} and C.N. Pope\hoch{\dagger,\ddagger}
}

\begin{document}

\begin{flushright}
\hfill{
MIFP-08-29}\\
\end{flushright}

\vspace{25pt}

\begin{center}

{\large {\bf Kerr-AdS/CFT Correspondence in Diverse Dimensions}}

\vspace{25pt}
\auth

\vspace{10pt}
\hoch{\dagger}{\tamphys}

\vspace{10pt}


\vspace{10pt}

\hoch{\ddagger}{\it  DAMTP, Centre for Mathematical Sciences,
 Cambridge University,\\  Wilberforce Road, Cambridge CB3 OWA, UK}

\vspace{40pt}

\underline{ABSTRACT}
\end{center}

   It was proposed recently that the near-horizon states of an
extremal four-dimensional Kerr black hole could be identified with a
certain chiral conformal field theory whose Virasoro algebra arises as
an asymptotic symmetry algebra of the near-horizon Kerr geometry.
Supportive evidence for the proposed duality came from the equality of
the microscopic entropy of the CFT, calculated by means of the Cardy
formula, and the Bekenstein-Hawking entropy of the extremal Kerr black
hole.  In this paper we examine the proposed Kerr/CFT correspondence
in a broader context.  In particular, we show that the microscopic
entropy and the Bekenstein-Hawking entropy agree also for the extremal
Kerr-AdS metric in four dimensions, and also for the extremal Kerr-AdS
metrics in dimensions 5, 6 and 7.  General formulae for all higher
dimensions are also presented.

\vspace{15pt}

\thispagestyle{empty}

\pagebreak
%
%


\section{Introduction}

   In a recent paper, it was shown that the near-horizon states of an
extremal four-dimensional Kerr black hole could be identified with a
certain two-dimensional chiral conformal field theory \cite{guhasost}.
This CFT arises by considering the asymptotic symmetry generators
associated with a class of perturbations around the near-horizon Kerr
geometry that obey suitably-chosen boundary conditions.  The Cardy
formula then gives a microscopic formula for the entropy for the CFT,
and it was shown that this coincides with the Bekenstein-Hawking
entropy of the extremal Kerr black hole.  This led the authors of
\cite{guhasost} to conjecture that the extreme Kerr black hole is
holographically dual to a chiral two-dimensional CFT whose central
charge is proportional to the angular momentum of the black hole.
(See \cite{carlip,sol1,par1,par2,hot} for some earlier related
work.)

   In this paper, we test the conjectured Kerr/CFT Correspondence for
a wider variety of extremal rotating black holes.  Firstly, we
consider the case of the Kerr-AdS black hole in four dimensions,
beginning with the construction of the near-horizon geometry of the
extremal Kerr-AdS black hole \cite{carter}.  We show that again in
this generalisation to include a cosmological constant, there is an
agreement between the Cardy formula for the CFT and the
Bekenstein-Hawking entropy of the extremal black hole.  We then turn
to the consideration of higher-dimensional rotating black
holes \cite{myper,glpp1,glpp2}, in the extremal limit,
beginning with a detailed discussion
of the five-dimensional case \cite{hawhuntay}.  In dimensions higher
than four, there is the new feature that there is more than one
angular momentum, since the black hole can rotate independently in
multiple mutually-orthogonal spatial 2-planes.  We now find that there
is a chiral two-dimensional CFT associated with each of these
rotations.  The central charge is different for each CFT, but
nevertheless, the Cardy formula gives rise to the same microscopic
entropy for the states in each CFT, and furthermore, this microscopic
entropy matches perfectly with the Bekenstein-Hawking entropy of the
extremal rotating black hole.  This agreement holds for both
Ricci-flat black holes and those with a cosmological constant.  Having
demonstrated this explicitly for the case of five-dimensional extremal
rotating black holes, we then present the analogous results for all
higher dimensions.  We have explicitly verified these expressions, and
hence the conjectured duality, in the cases of six and seven
dimensional extremal black holes.  Our results include a construction
of the metrics describing the near-horizon geometries for extremal
Kerr-AdS black holes in all dimensions.  

\section{Four-Dimensional Extremal Kerr-AdS Black Holes}

\subsection{The near-horizon metric}

   The metric of the four-dimensional Kerr-AdS black hole, satisfying
$R_{\mu\nu}=- 3\ell^{-2}\, g_{\mu\nu}$, is given by
\cite{carter}
\bea
ds^2&=& \rho^2 \Big(\fft{d\hat r^2}{\Delta} + \fft{d\theta^2}{\Delta_{\theta}}
\Big) + \fft{\Delta_\theta \sin^2\theta}{\rho^2} \Big(a d\hat t -
\fft{\hat r^2 + a^2}{\Xi} d\hat \phi\Big)^2 -
\fft{\Delta}{\rho^2} (d\hat t - \fft{a \sin^2\theta}{\Xi} d\hat \phi
\Big)^2\,,\nn\\
\rho^2&=& \hat r^2 + a^2 \cos^2\theta\,,\qquad
\Delta=(\hat r^2 + a^2) (1 + \hat r^2\ell^{-2}) - 2 M \hat r\,,\nn\\
&&
\Delta_\theta = 1 - a^2\ell^{-2} \cos^2\theta\,,\qquad
\Xi=1 - a^2\ell^{-2}\,.\label{d4kerr}
\eea
The outer horizon is located at $r=r_+$, the largest root of
$\Delta=0$.  Note that in this coordinate system the metric is
asymptotic to AdS$_4$ in a rotating frame, with angular velocity
$\Omega_{\infty}=-a \ell^{-2}$.

   The Hawking temperature, the Bekenstein-Hawking entropy, the
angular momentum and the angular velocity of the horizon (as measured
in the asymptotically rotating frame) are given by
\bea
T_H&=&\fft{r_+^2 -a^2 + r_+^2 \ell^{-2} (3r_+^2 + a^2)}{4\pi r_+
(r_+^2 + a^2)} \,,\qquad S_{BH} = \fft{\pi (r_+^2+ a^2)}{\Xi}\,,\nn\\
J_\phi &=&  \fft{M\,a}{\Xi^2}\,,\qquad
\Omega_\phi =\fft{a\, \Xi}{r_+^2 + a^2}\,.\label{d4thermo}
\eea

   The extremal limit is attained when the function $\Delta$ has a
double zero at the outer horizon, which we shall denote by $r=r_0$,
\ie when $\Delta(r_0)=0$ and $\Delta'(r_0)=0$, implying that the
Hawking temperature vanishes.  In order to avoid expressions
explicitly involving roots of the polynomial equation, it is
convenient to use these two conditions in order to express the mass
parameter $M$ and the rotation parameter $a$ in terms of $r_0$:
\be
M=\fft{r_0 (1 + r_0^2\ell^{-2})^2}{1 - r_0^2\ell^{-2}}\,,\qquad
a^2=\fft{r_0^2(1 + 3 r_0^2\ell^{-2})}{1-r_0^2\ell^{-2}}\,.
\ee

   Near the horizon of the extremal black hole, 
the function  $\Delta$ takes the form
\be
\Delta = (\hat r-r_0)^2\, V + {\cal O}\Big( (\hat r-r_0)^3 \Big)\,,\qquad
\hbox{with}\qquad
V=\fft{1+6r_0^2\ell^{-2} - 3 r_0^4\ell^{-4}}{1 -r_0^2\ell^{-2}}\,.
\ee
To obtain the near-horizon geometry, we make the coordinate
transformations
\be
\hat r = r_0 (1 + \lambda y)\,,\qquad
\hat\phi = \phi + \fft{a\,\Xi}{r_0^2 + a^2} \hat t\,,\label{d4scale1}
\ee
and scale the time $\hat t$ by writing
\be
\hat t = \fft{r_0^2 + a^2}{\lambda r_0 V}t\,.\label{d4scale2}
\ee
Sending $\lambda\rightarrow 0$ then gives the near-horizon metric
\bea
ds^2 &=& \fft{\rho_0^2}{V}\Big(-y^2 dt^2 + \fft{dy^2}{y^2} +
\fft{Vd\theta^2}{\Delta_\theta}\Big)
+\fft{\sin^2\theta\, \Delta_\theta}{\rho_0^2}
\Big(\fft{2ar_0}{V} y dt + \fft{r_0^2 + a^2}{\Xi} d\phi\Big)^2\,,\nn\\
\rho_0^2&=&r_0^2 + a^2\cos^2\theta\,.\label{d4nhek1}
\eea

     Finally, we can transform the AdS$_2$ Poincar\'e coordinates
$y$ and $t$ to global coordinates $r$ and $\tau$ as follows.  AdS$_2$ is
described by the hyperboloid $Z^2-X^2-Y^2=-1$ in $\R^3$, whose
parameterisations in the Poincar\'e and global coordinate systems are
\bea
\hbox{Poincar\'e}:&&
 X+Z=y\,,\qquad X-Z= \fft1{y} - y t^2\,,\qquad Y=y\, t\,,\nn\\
\hbox{Global}:&& X= \sqrt{1+r^2}\, \cos\tau\,,\qquad Y=\sqrt{1+r^2}\,
\sin\tau
\,,\qquad Z =r\,.
\eea
 From this, we see that
\be
y= r+ \sqrt{1+r^2}\, \cos\tau\,,\qquad
 t= \fft{\sqrt{1+r^2}\, \sin\tau}{r+ \sqrt{1+r^2}\, \cos\tau}\,,
\label{poinglob}
\ee
and hence we find that
\bea
-y^2\, dt^2 + \fft{dy^2}{y^2} &=& -(1+r^2)\, d\tau^2 + \fft{dr^2}{1+r^2}
\,,\nn\\
y \, dt &=& r\, d\tau + d\gamma\,,
\eea
where
\be
\gamma\equiv
  \log\Big(\fft{1+ \sqrt{1+r^2}\, \sin\tau}{\cos \tau +
r\, \sin\tau}\Big)\,.\label{d2gamma}
\ee
Thus by sending
\be
\phi\longrightarrow \phi  +
\fft{2 a r_0\, \Xi\, \gamma}{(r_0^2+a^2)\, V}\,,
\ee
the near-horizon metric (\ref{d4nhek1}) becomes
\be
ds^2 = \fft{\rho_0^2}{V}\Big(-(1+r^2) d\tau^2 + \fft{dr^2}{1+r^2} +
\fft{Vd\theta^2}{\Delta_\theta}\Big)
+\fft{\sin^2\theta\, \Delta_\theta}{\rho_0^2}
\Big(\fft{2ar_0}{V} r d\tau + \fft{r_0^2 + a^2}{\Xi} d\phi\Big)^2\,,
\label{d4nhek2}
\ee
In the case of vanishing cosmological constant, the metric becomes
the near-horizon geometry of the Kerr solution, analysed in
\cite{barhor}.

\subsection{Central charge in the Virasoro algebra}\label{d4centsec}

  Here we follow the procedures described in \cite{guhasost}, in which
the asymptotic symmetry algebra for a certain class of diffeomorphisms
of the near-horizon extremal metric is studied.  A crucial aspect of
these diffeomorphisms is that the deviations $h_{\mu\nu}$ of the
background metric are allowed, in the $h_{\tau\tau}$ and
$h_{\phi\phi}$ directions, to be of the same order in inverse powers
of $r$ as the corresponding components in the background metric
itself.  This is less restrictive than the asymptotically AdS boundary
conditions considered in \cite{browhenn}, where all metric components
are required to be at sub-leading order.  We refer the reader to
\cite{guhasost} for a detailed discussion of this point. By allowing
the more general boundary conditions, it was shown in \cite{guhasost}
that a class of diffeomorphisms around the near-horizon extremal Kerr
metric gave rise to a chiral Virasoro algebra.  We now apply the
analogous analysis to the case with a cosmological constant.

   The relevant diffeomorphisms are given by the vector fields
\cite{guhasost}
\be
\zeta_\ep = \ep(\phi)\, \fft{\del}{\del\phi} - 
   r\, \ep'(\phi)\, \fft{\del}{\del r}\,.\label{d4diffeo}
\ee
The function $\ep(\phi)$ is periodic in the azimuthal angle, and we may
consider a mode expansion in terms of diffeomorphisms $\zeta_\n$ in which
we take $\ep(\phi) = - e^{-\im n\phi}$:
\be 
\zeta_\n = - e^{-\im n \phi}\, \fft{\del}{\del\phi} - \im n\, r\, 
   e^{-\im n \phi}\, \fft{\del}{\del r}\,.
\ee
The commutator of these vector fields generates the centreless Virasoro
algebra,
\be
\im \, [ \zeta_\m, \zeta\n] = (m-n) \zeta_{\sst{(m+n)}}\,.
\ee

   Following the discussion in \cite{guhasost}, charges $Q_\zeta$
associated with these diffeomorphisms are defined by integrals over a
spatial slice $\del\Sigma$,
\be
Q_\zeta = \fft1{8\pi}\, \int_{\del\Sigma} k_\zeta\,,\label{Qdefs}
\ee
where the 2-form $k_\zeta$ is defined for a perturbation $h_{\mu\nu}$
around the background metric $g_{\mu\nu}$ 
by
\bea
k_\zeta[h,g] &=&\ft12 \Big[ \zeta_\nu\nabla_\mu h
  - \zeta_\nu \nabla_\sigma h_\mu{}^\sigma +
  \zeta_\sigma\nabla_\nu h_\mu{}^\sigma + \ft12 h \nabla_\nu\zeta_\mu
- h_\nu{}^\sigma \nabla_\sigma\zeta_\mu\nn\\
&&\qquad\qquad\qquad\ft12 h_{\nu\sigma}
(\nabla_\mu\zeta^\sigma + \nabla_\sigma\zeta_\mu)\Big] \, {*(dx^\mu\wedge
dx^\nu)}\,,\label{kdef}
\eea
where $*$ denotes the Hodge dual.  Note that the formula also
applies to higher dimensions. 
The Dirac bracket algebra of the charges includes a central term
\cite{barbra}:
\be
\{Q_{\zeta_\m},Q_{\zeta_\n}\}_{DB} = Q_{[\zeta_\m,\zeta_\n]} +
  \fft1{8\pi}\, \int_{\del\Sigma} k_{\zeta_\m}[{\cal L}_{\zeta_\n}g,g]\,,
\label{centralterm}
\ee
where ${\cal L}_\zeta g_{\mu\nu}=
  \zeta^\rho\del_\rho g_{\mu\nu} + g_{\rho\nu}\, \del_\mu \zeta^\rho +
g_{\mu\rho}\, \del_\nu \zeta^\rho$ is the Lie derivative of $g_{\mu\nu}$
with respect to $\zeta$. Translated into modes, the corresponding
commutator algebra is given by
\be
[L_m,L_n]= (m-n) L_{m+n} + \fft1{12} c_L\, (m^3 + \alpha m)\delta_{m+n,0}\,,
\ee
with the central charge $c_L$ given by
\be
\fft1{8\pi}\, \int_{\del\Sigma} k_{\zeta_\m}[{\cal L}_{\zeta_\n}g,g]=
 -\fft{\im}{12} c_L\, (m^3 + \alpha m)\, \delta_{m+n,0}\,.\label{central}
\ee
(The term linear in $m$ plays no essential r\^ole here, since it can
be shifted arbitrarily by $c$-number shifts of the generators $L_m$.
Thus the value of the constant $\alpha$ is unimportant.)

   Evaluating the integral (\ref{central}) for the case of the
four-dimensional near-horizon extremal Kerr-AdS metric
(\ref{d4nhek2}), we find
\bea
\fft1{8\pi}\, \int_{\del\Sigma} k_{\zeta_\m}[{\cal L}_{\zeta_\n}g,g]
&=&
 -\fft{\im m^3 r_0^2 \sqrt{(1-r_0^2\, \ell^{-2})(1+3r_0^2\, \ell^{-2})}}{
1 + 6r_0^2\, \ell^{-2} - 3 r_0^4\, \ell^{-4}}\nn\\
&&-
  \fft{2\im m r_0^2 \sqrt{(1-r_0^2\, \ell^{-2})(1+3r_0^2\, \ell^{-2})}}{
  (1- 3 r_0^2 \, \ell^{-2})^2}\,,
\eea
and hence the central charge is
\be
c_L=\fft{12r_0^2 \sqrt{(1-r_0^2\, \ell^{-2})(1+3r_0^2\, \ell^{-2})}}{
1 + 6r_0^2\, \ell^{-2} - 3 r_0^4\, \ell^{-4}}\,.\label{d4cent}
\ee

\subsection{Microscopic entropy and the Cardy formula}\label{d4cardysec}

   Following \cite{guhasost}, one can adopt the Frolov-Thorne vacuum
\cite{frotho} to provide a definition of the vacuum state for the
extreme Kerr-AdS metric.  Quantum fields for the general
(non-extremal) Kerr-AdS metric (\ref{d4kerr}) can be expanded in
eigenstates with asymptotic energy $\omega$ and angular momentum $m$,
with $\hat t$ and $\hat\phi$ dependence $e^{-\im \omega\hat t+ \im
m\hat\phi}$.  In terms of the the redefined $t$ and $\phi$ coordinates
of the extremal near-horizon limit, given by (\ref{d4scale1}) and
(\ref{d4scale2}), we have
\be 
e^{-\im \omega\hat t+ \im m\hat\phi} = e^{-\im n_R t + \im n_L \phi}\,,
\ee
where
\be 
n_L = m\,,\qquad n_R = \fft{(r_0^2+a^2)\omega - a m\, \Xi}{\lambda r_0\, V}
\,.
\ee
The left-moving and right-moving temperatures $T_L$ and $T_R$ are then
defined by writing the Boltzmann factor $e^{-(\omega - m
\Omega_\phi)/T_H}$ as
\be 
e^{-(\omega - m \Omega_\phi)/T_H}= e^{-n_L/T_L - n_R/T_R}\,.
\ee
This allows us to read off
\be
T_L= \fft{T_H}{a\, \Xi\,(r_0^2+a^2)^{-1}  -  \Omega_\phi}\,,\qquad
T_R = \fft{(r_0^2+a^2)\, T_H}{\lambda r_0\, V}\,.\label{d4TLTR}
\ee
We now take the extremal limit, finding\footnote{Note that had we
chosen to work in coordinate frame in which the angular velocity at
infinity were different (such as an asymptotically static frame), then
the expressions for $\Omega_\phi$ and for the $\hat t$-dependent shift
in (\ref{d4scale1}) would each be be displaced by the same additive
constant, and so the final expression for $T_L$ in (\ref{d4TLTR})
would be the same.}
\be 
T_L= \fft{1+6r_0^2\, \ell^{-2} - 3r_0^4\, \ell^{-4}}{2\pi
(1-3r_0^2\, \ell^{-2})\sqrt{(1 + 3r_0^2\, \ell^{-2}) (1-r_0^2\,
\ell^{-2})}}\,,\qquad 
T_R= 0\,.\label{d4T}
\ee
In the case of vanishing cosmological constant, {\it i.e.} $\ell^{-1}=0$,
the result reduces to $T_L=1/(2\pi)$, given in \cite{guhasost}.

   From the Cardy formula for the entropy of a unitary conformal field
theory at temperature $T_L$, the microscopic entropy is given by
\be 
S= \ft13 \pi^2\,  c_L\, T_L\,.
\ee
 From (\ref{d4cent}) and (\ref{d4T}), we therefore obtain the microscopic
entropy 
\be
S= \fft{2\pi r_0^2}{1 - 3 r_0^2\, \ell^{-2}}
\ee
for the CFT dual to the extremal Kerr-AdS black hole.  This agrees
perfectly with the Bekenstein-Hawking entropy which given in
(\ref{d4thermo}), after taking the extremal limit.

\section{Five-Dimensional Extremal Rotating Black Holes}

\subsection{The extremal near-horizon metric}

   The general result for the metric of a five-dimensional rotating
black hole with $S^3$ horizon topology, satisfying the Einstein
equation $R_{\mu\nu}= -4\ell^{-2}\, g_{\mu\nu}$, was obtained by
Hawking, Hunter and Taylor-Robinson \cite{hawhuntay}. This metric,
which generalises the Ricci-flat rotating black hole of Myers and
Perry \cite{myper}, is given by
\bea
ds^2&=& - \fft{\Delta}{\rho^2} (d\hat t - \fft{a\sin^2\theta}{\Xi_a}
d\phi_1 - \fft{b\cos^2\theta}{\Xi_b} d\phi_2)^2 +
\fft{\Delta_\theta \sin^2\theta}{\rho^2}
(a d\hat t - \fft{(\hat r^2 + a^2)}{\Xi_a} d\phi_1)^2\nn\\
&&+\fft{\Delta_\theta \cos^2\theta}{\rho^2} (b d\hat t -
\fft{(\hat r^2 + b^2)}{\Xi_b} d\phi_2)^2 +
\fft{\rho^2}{\Delta} d\hat r^2 + \fft{\rho^2}{\Delta\theta} d\theta^2\\
&& +\fft{1 + \hat r^2\ell^{-2}}{\hat r^2\rho^2}
\Big(ab d\hat t - \fft{b(\hat r^2 + a^2) \sin^2\theta}{\Xi_a}
d\phi_1 - \fft{a(\hat r^2 + b^2)\cos^2\theta}{\Xi_b} d\phi_2\Big)^2
\,,\nn
\eea
where
\bea
&&\Delta = \fft{1}{\hat r^2}(\hat r^2 + a^2)(\hat r^2 + b^2)
(1 + \hat r^2 \ell^{-2}) - 2M\,,\quad
\Delta_\theta=1 - a^2\ell^{-2} \cos^2\theta - b^2\ell^{-2} \sin^2\theta
\,,\nn\\
&&\rho^2 = \hat r^2 + a^2\cos^2\theta + b^2\sin^2\theta\,,\qquad
\Xi_a=1-a^2\ell^{-2}\,,\qquad
\Xi_b=1 - b^2 \ell^{-2}\,.
\eea
Note that in this coordinate system, the metric is asymptotic to
AdS$_5$ in a rotating frame, with angular velocities
$\Omega_{\phi_1}^\infty=-a \ell^{-2}$ and $\Omega_{\phi_2}^\infty=-b
\ell^{-2}$.

   The Hawking temperature, Bekenstein-Hawking entropy, the
angular momenta and  the angular 
velocities on the horizon (in the asymptotically rotating frame) 
are given by
\bea
T_H&=& \fft{r_+^4 - a^2 b^2 + g^2 r_+^4 (2r_+^2 + a^2 + b^2)}{2\pi
r_+ (r_+^2 + a^2)(r_+^2 + b^2)}\,,\nn\\
S_{BH} &=& \fft{\pi^2 (r_+^2 + a^2) (r_+^2 + b^2)}{2 r_+ \Xi_a\Xi_b}\nn\\
J_{\phi_1}&=& \fft{\pi M a}{2\Xi_1^2 \Xi_2}\,,\qquad
J_{\phi_2}= \fft{\pi M b}{2\Xi_1\Xi_2^2}\,,\nn\\
\Omega_{\phi_1} &=& \fft{a\, \Xi_a}{r_+^2 + a^2}\,,
\qquad \Omega_{\phi_2} = \fft{b\, \Xi_b}{r_+^2 + b^2}\,.
\label{d5therm}
\eea

    Extremality occurs if $\Delta$ has a double zero at the horizon radius
$r=r_0$, implying that the Hawking temperature vanishes.  These conditions
can conveniently be solved for $M$ and $\ell^{-2}$: 
\be
M=\fft{(r_0^2 + a^2)^2 (r_0^2 + b^2)^2}{2r_0^4 (2r_0^2 + a^2 + b^2)}\,,
\qquad
\fft{(a b -r_0^2)(a b + r_0^2)}{r_0^4 (2r_0^2 + a^2 + b^2)} = \ell^{-2}\,.
\label{d5extreme}
\ee
With these parameters, we have $\Delta = (r-r_0)^2 V + {\cal O}((r-r_0)^3)$,
where
\be
V=4 + 4(3r_0^2 + a^2 + b^2)\ell^{-2}\,.
\ee
To extract the near-horizon
geometry we first make the coordinate transformations
\bea
&&\hat r=r_0 (1 + \lambda y)\,,\qquad
\phi_1\rightarrow \phi_1 + \alpha_1\, \hat t\,,\qquad
\phi_2\rightarrow \phi_2 + \alpha_2\, \hat t\,,\nn\\
&&
\alpha_1= \fft{a\,\Xi_a}{r_0^2 + a^2}\,,\qquad
\alpha_2= \fft{b\,\Xi_b}{r_0^2 + b^2}\,.\label{d5alphas}
\eea
then scale the time coordinate
\be
\hat t=\beta\, t\,,\qquad
\beta=\fft{(r_0^2 + a^2)(r_0^2 + b^2)}{\lambda\, r_0^3 V}\,,
\label{d5beta}
\ee
and finally send $\lambda \rightarrow 0$.  The metric becomes
\bea
ds^2&=& \fft{\rho_0^2}{V} \Big (-y^2 dt + \fft{dy^2}{y^2}\Big) +
\fft{\rho_0^2 d\theta^2}{\Delta_\theta}+
\fft{\Delta_\theta \sin^2\theta}{\rho_0^2}
\Big(\fft{2a(r_0^2 + b^2)}{r_0 V} y dt + \fft{r_0^2 + a^2}{\Xi_a}
d\phi_1\Big)^2
\nn\\
&&+
\fft{\Delta_\theta\cos^2\theta}{\rho_0^2}\Big(
\fft{2b(r_0^2 + a^2)}{r_0 V} y dt + \fft{r_0^2 + b^2}{\Xi_b} d\phi_2
\Big)^2 \label{d5nhek1}\\
&&+\fft{1 + r_0^2 \ell^{-1}}{r_0^2 \rho_0^2} \Big(
\fft{2 a b \rho_0^2}{r_0 V} y dt + \fft{b (r_0^2 + a^2)\sin^2\theta}{\Xi_a}
d\phi_1 + \fft{a(r_0^2 + b^2)\cos^2\theta}{\Xi_b} d\phi_2\Big)^2\,,\nn
\eea
where
\be
\rho_0^2=r_0^2 + a^2 \cos^2\theta + b^2\sin^2\theta\,,\qquad
\Delta_\theta = 1 - a^2\ell^{-2} \cos^2\theta - b^2 \ell^{-2} \sin^2\theta
\,.
\ee

   Finally, we can transform the AdS$_2$ Poincar\'e coordinates $y$
and $t$ to global coordinates $r$ and $\tau$ by using the
transformations (\ref{poinglob}), and sending
\be
\phi_1\longrightarrow \phi_1 + \fft{2a\gamma\, (r_0^2+b^2)\, \Xi_a}{
      r_0\, V\, (r_0^2+a^2)}\,,\quad
\phi_2\longrightarrow \phi_2 + \fft{2b\gamma\, (r_0^2+a^2)\, \Xi_b}{
      r_0\, V\, (r_0^2+b^2)}\,,
\ee
where $\gamma$ is given by (\ref{d2gamma}).
The five-dimensional near-horizon extremal black hole metric 
(\ref{d5nhek1}) becomes
{\crampest
\bea
ds^2&=& \fft{\rho_0^2}{V} \Big (-(1+r^2)\, d\tau^2 +
\fft{dr^2}{1+r^2}\Big) +
\fft{\rho_0^2 d\theta^2}{\Delta_\theta}+
\fft{\Delta_\theta \sin^2\theta}{\rho_0^2}
\Big(\fft{2a(r_0^2 + b^2)}{r_0 V} \, r\, d\tau +
\fft{r_0^2 + a^2}{\Xi_a} d\phi_1\Big)^2
\nn\\
&&+
\fft{\Delta_\theta\cos^2\theta}{\rho_0^2}\Big(
\fft{2b(r_0^2 + a^2)}{r_0 V} \, r\, d\tau +
\fft{r_0^2 + b^2}{\Xi_b} d\phi_2
\Big)^2 \label{d5nhek2} \\
&&+\fft{1 + r_0^2 \ell^{-2}}{r_0^2 \rho_0^2} \Big(
\fft{2 a b \rho_0}{r_0 V}\, r\, d\tau +
\fft{b (r_0^2 + a^2)\sin^2\theta}{\Xi_a}
d\phi_1 + \fft{a(r_0^2 + b^2)\cos^2\theta}{\Xi_b}
d\phi_2\Big)^2\,.\nn
\eea
}
(This result also appears, with $\ell^{-1}=0$, in \cite{kun1}, and
with $\ell^{-1} >0$ in \cite{kun2}.)

\subsection{Central charge and microscopic entropy}

   We can introduce an asymptotic symmetry group for this
five-dimensional extremal black hole in a manner that closely
parallels the four-dimensional example in \cite{guhasost}.  The
difference now is that instead of having just a single algebra of
reparameterisations of the circle, as in (\ref{d4diffeo}), there are
now two, associated with the two independent azimuthal coordinates
$\phi_1$ and $\phi_2$:
\bea
\zeta^\1_\n = - e^{-\im n \phi_1}\, \fft{\del}{\del\phi_1} - \im n\, r\,
   e^{-\im n \phi_1}\, \fft{\del}{\del r}\,,\nn\\
\zeta^\2_\n = - e^{-\im n \phi_2}\, \fft{\del}{\del\phi_2} - \im n\, r\,
   e^{-\im n \phi_2}\, \fft{\del}{\del r}\,,
\eea

   We can now calculate the central charges for each of the
independent (mutually commuting) Virasoro algebras, using the
formalism summarised in section \ref{d4centsec}.  We find
\bea
c_{\phi_1}&=&\fft{6\pi a (r_0^2 + b^2)^2}{r_0^2 \Xi_b V}=
\fft{3\pi a r_0^6(r_0^2 + b^2) (2r_0^2 + a^2 + b^2)^2}{
2(2r_0^4 + a^2 r_0^2 -a^2b^2)(-r_0^6 + 3a^2b^2 r_0^2 +
a^2b^4+a^4b^2)}\,,\nn\\
c_{\phi_2} &=&\fft{6\pi b(r_0^2 + a^2)^2}{r_0^2 \Xi_a V} =
\fft{3\pi b r_0^6(r_0^2 + a^2)(2r_0^2 + a^2 + b^2)^2}{
2(2r_0^4 + b^2 r_0^2 -a^2b^2)(-r_0^6 + 3a^2b^2 r_0^2 +
a^2b^4 + a^4 b^2)}\,.\label{d5cent}
\eea
 
  There are now three Frolov-Thorne temperatures associated with the
quantum field theory of the extremal five-dimensional Kerr-AdS metric,
which we shall denote by $T_t$, $T_{\phi_1}$ and $T_{\phi_2}$.  Thus
we consider
\be
e^{-{\rm i} \omega \,\hat t + {\rm i} m_1\, \hat \phi_1 + {\rm i} m_2\,
\hat \phi_2} =e^{-{\rm i} n_t \,t + {\rm i} n_{\phi_1} \,\phi_1 +
{\rm i} n_{\phi_2}\, \phi_2}\,,
\ee
where
\be
\hat\phi_1 = \phi_1 + \alpha_1\, \hat t\,,\qquad
\hat\phi_2 = \phi_2 + \alpha_2\, \hat t\,,\qquad
\hat t=\beta\, t\,.
\ee
This implies that
\be
n_{\phi_1} = m_1\,,\qquad n_{\phi_2} = m_2\,,\qquad
\omega=\fft{n_t}{\beta} + \alpha_1\, n_{\phi_1} +
\alpha_2\, n_{\phi_2}\,.
\ee
Now we consider
\be
\exp\Big(-\fft{\omega - m_1 \Omega_{\phi_1} - m_2 \Omega_{\phi_2}}{T_H}\Big)
=\exp\Big(-\fft{n_t}{T_t} - \fft{n_{\phi_1}}{T_{\phi_1}} -
\fft{n_{\phi_2}}{T_{\phi_2}}\Big)\,.
\ee
Thus we have
\be
T_t=\beta T\,,\qquad T_{\phi_1} = \fft{T_H}{\alpha_1-\Omega_1}
\,,\qquad T_{\phi_2} = \fft{T_H}{\alpha_2 - \Omega_2}\,.
\ee
where $\alpha_1, \alpha_2$ and $\beta$ are given in (\ref{d5alphas})
and (\ref{d5beta}).
In the extremal limit, these quantities evaluate to give
\bea
T_{t}&=&0\,,\nn\\
T_{\phi_1}&=&\fft{r_0 (r_0^2 + a^2) V}{4\pi a \Xi_a (r_0^2 + b^2)}
=\fft{r_0(-r_0^6 + 3a^2 b^2 r_0^2 + a^2 b^4 + a^4 b^2)}{
\pi a (r_0^2 + b^2) (2 r_0^4 + b^2 r_0^2 - a^2 b^2)}\,,\nn\\
T_{\phi_2}&=&\fft{r_0 (r_0^2 + b^2)V}{4\pi b \Xi_b (r_0^2 + a^2)}=
\fft{r_0(-r_0^6+ 3a^2 b^2 r_0^2 +a^2 b^4 + a^4 b^2)}{
\pi b (r_0^2 +a^2)(2r_0^4+a^2r_0^2 - a^2 b^2)}\,.\label{d5frolov}
\eea

   We can use the Cardy formula to calculate the microscopic entropy
of the CFT associated with each of the $\phi_1$ and $\phi_2$ circles.
The central charges (\ref{d5cent}) and Frolov-Thorne temperatures
(\ref{d5frolov}) are different for the two CFTs, but nevertheless we
find that the microscopic entropy of each CFT is the same:
\be
S=\ft13 \pi^2 c_{\phi_1} T_{\phi_1}=
\ft13 \pi^2 c_{\phi_2} T_{\phi_2}
\ee
Furthermore, this entropy agrees precisely with the Bekenstein-Hawking
entropy of the extremal five-dimensional black hole, which is obtained
by substituting the extremality 
conditions (\ref{d5extreme}) into (\ref{d5therm}): 
\be
S_{BH}=\fft{\pi^2 r_0^7 (2r_0^2 + a^2+b^2)^2}{2(2r_0^4 + a^2 r_0^2 - a^2b^2)
(2r_0^4 + b^2 r_0^2 - a^2 b^2)}\,.
\ee
In the Ricci-flat case $\ell^{-1}=0$, which implies $r_0^2=a\,b$,
we have the much simpler expressions
\bea
&&T_{\phi_1}=\fft{a}{\pi\sqrt{a b}},\qquad
T_{\phi_2}=\fft{b}{\pi\sqrt{a b}}\,,\qquad
c_{\phi_1}=\ft32 \pi b(a+b)^2\,,\qquad
c_{\phi_2}=\ft32 \pi a(a+b)^2\,,\nn\\
&&S=\ft12\pi^2 (a+b)^2 \sqrt{ab}\,.
\eea

\section{Kerr-AdS/CFT Correspondence in Higher Dimensions}

The general Kerr-AdS metrics in arbitrary higher dimensions were
obtained in \cite{glpp1,glpp2}.  The solutions were later generalised
further to include NUT charges \cite{chlupo}.  Although we shall
not be including the NUT parameters in our calculations, the form of
the metrics obtained in \cite{chlupo} will be the most convenient one
for our purposes.  We shall first review the Kerr-AdS metrics written
in this form, and then obtain their extremal limits.  It is convenient
to consider the cases when the spacetime dimension is odd, $D=2n+1$,
and even, $D=2n$, separately.

\subsection{$D=2n+1$ dimensional Kerr-AdS black holes,
and extremal limit}

  We shall use the coordinates introduced in section 2 of
\cite{chlupo}, except that here we shall shift the azimuthal angles by
adding convenient constant multiples of the time coordinate $\hat t$.
Although the resulting coordinate frame is asymptotically rotating, it
has the advantage of simplifying the expression for the metric.  (As
we discussed in section 2.3 for the four-dimensional example, the
final results are independent of the choice of frame.)  The metric is
then given by
\bea
ds^2 &=& \fft{U}{X} d\hat r^2 + \sum_{\alpha=1}^{n-1}
\fft{U_\alpha}{X_\alpha} dy_\alpha^2 - \fft{X}{U}
\Big[d\hat t - \sum_{i=1}^n a_i^2 \gamma_i
\fft{d\hat \phi_i}{\epsilon_i}\Big]^2\label{oddmet}\\
&&+\sum_{\alpha=1}^{n-1}\fft{X_\alpha}{U_\alpha}\Big[d\hat t -
\sum_{i=1}^n \fft{a_i^2(\hat r^2 + a_i^2)\gamma_i}{a_i^2-y_\alpha^2}
\fft{d\hat \phi_i}{\epsilon_i}\Big]^2
+\fft{\prod_{k=1}^n a_k^2}{\hat r^2 \prod_{\alpha=1}^{n-1} y_\alpha^2}
\Big[d\hat t - \sum_{i=1}^n (\hat r^2 + a_i^2) \gamma_i
\fft{d\hat\phi_i}{\epsilon_i}\Big]^2\,,\nn
\eea
where
\bea
U&=&\prod_{\alpha=1}^{n-1} (\hat r^2 + y_\alpha^2)\,,\qquad
U_\alpha=-(\hat r^2 + y_\alpha^2) {\prod}_{\beta=1}^{'n-1}
(y_\beta^2-y_\alpha^2)\,,\quad
1\le\alpha\le n-1\,,\nn\\
\epsilon_i &=& a_i \Xi_i {\prod}^{'n}_{k=1}
(a_i^2-a_k^2)\,,\qquad
\gamma_i=\prod_{\alpha=1}^{n-1} (a_i^2-y_\alpha^2)\,,\quad
1\le i\le n\,,\nn\\
X&=&\fft{1 + \hat r^2\ell^{-2}}{\hat r^2}
\prod_{k=1}^n (\hat r^2 + a_k^2) - 2 M\,,\qquad
\Xi_i =1 - a_i^2 \ell^{-2}\,,\nn\\
X_\alpha&=& \fft{1-y_\alpha^2\ell^{-2}}{y_\alpha^2}
\prod_{k=1}^n (a_k^2 - y_\alpha^2) + 2 L_\alpha\,,
\qquad 1\le\alpha\le n-1\,.
\label{functionsodd}
\eea
Note that the notation $\prod'$ indicates that the term in
the full product that vanishes is to be omitted.
We shall focus on the Kerr-AdS solutions, corresponding to setting all
the NUT charges zero, {\it i.e.} $L_\alpha=0$ for all $1\le \alpha\le
n-1$.  Without loss of generality, we may order the rotation
parameters such that $a_1\le a_2\le \cdots \le a_n$, in which case the
compact coordinates $y_\alpha$ must lie in the ranges $a_\alpha \le
y_\alpha \le a_{\alpha +1}$.  The asymptotic region, where the metric
approaches AdS$_{2n+1}$ written in global coordinates, is at $\hat
r=\infty$.  The $n$ azimuthal angles $\hat \phi_i$ each have period
$2\pi$, and the asymptotic angular velocities are $\Omega_i^\infty=-a_i
\ell^{-2}$.

      The horizon is located at $\hat r=r_+$, where $r_+$ is the
largest root of the polynomial function $X$.  The thermodynamic
quantities for the Kerr-AdS solutions were obtained in \cite{gipepo}.
The mass, angular momenta and Bekenstein-Hawking entropy are given by
\be
E=\fft{m {\cal A}_{D-2}}{4\pi (\prod_j \Xi_j)}
\Big(\sum_{i=1}^n\fft{1}{\Xi_i} - \ft12\Big)\,,\qquad
J_i= \fft{M a_i {\cal A}_{D-2}}{4\pi \Xi_i (\prod_j \Xi_j)}\,,\qquad
S_{BH}=\fft{{\cal A}_{D-2}}{4\pi r_+} \prod_{i=1}^n \fft{r_+^2 + a_i^2}{\Xi_i}
\,,\label{oddthermo}
\ee
where ${\cal A}_{D-2} = 2\pi^{(D-1)/2}/\Gamma[(D-1)/2]$ is the
volume of the unit $(D-2)$-sphere.
 The Hawking temperature and the angular velocities on the horizon
(measured in the asymptotically rotating frame) are given by
\bea
T_H&=&\fft1{2\pi}\, \Big[
r_+(1+ r_+^2\ell^{-2}) \sum_i \fft{1}{r_+^2 + a_i^2} -
\fft{1}{r_+}\Big] \,,\nn\\
\Omega_i&=& \fft{a_i\Xi_i}{r_+^2 + a_i^2}\,.
\eea
      
   We now consider the extremal limit.  It occurs when the function
$X$ has a double zero, implying that the parameters must be chosen so
that
\be
X|_{\hat r=r_0}=0\,,\qquad X'|_{\hat r=r_0} = 0\,,
\ee
where $\hat r=r_0$ is the horizon of the extremal black hole.  Near
the horizon, the function $X$ can be expanded as
\be
X=(\hat r-r_0)^2 V + {\cal O}\Big( (\hat r-r_0)^3\Big)\,,
\ee
where
\be
V=\ft12 X''|_{\hat r=r_0}\,.
\ee

   To extract the near-horizon geometry, we first make the coordinate
redefinition
\be
\hat r = r_0 (1 + \lambda y)\,,\qquad
\hat \phi_i=\phi_i + \alpha_i \hat t\,,\qquad
\alpha_i=\fft{a_i \Xi_i}{r_0^2 + a_i^2}\,,
\label{alphaidef}
\ee
and then make a scaling of the time coordinate
\be
\hat t =\beta t\,,\qquad
\beta=\fft{\prod_i (r_0^2 + a_i^2)}{\lambda r_0^3 V}\,.
\label{betadef}
\ee
Finally, taking the limit $\lambda\rightarrow 0$ we obtain the
near-horizon geometry
\bea
ds^2 &=& \fft{\widetilde U}{V} \Big(-y^2 dt^2 + \fft{dy^2}{y^2}\Big)
+\sum_{\alpha=1}^{n-1}
\fft{\widetilde U_\alpha}{X_\alpha} dy_\alpha^2\\
&&+\sum_{\alpha=1}^{n-1}\fft{X_\alpha}{\widetilde U_\alpha}
\Big [\fft{2r_0\prod_{\beta} (r_0^2 + y_\beta^2)}{V(r_0^2 + y_\alpha^2)}
y dt +\sum_{i=1}^n \fft{a_i^2( r_0^2 + a_i^2)\gamma_i}{a_i^2-y_\alpha^2}
\fft{d\phi_i}{\epsilon_i}\Big]^2\nn\\
&&
+\fft{\prod_{k=1}^n a_k^2}{r_0^2 \prod_{\alpha=1}^{n-1} y_\alpha^2}
\Big[\fft{2\prod_\beta (r_0^2 + y_\beta^2)}{r_0 V}
y dt + \sum_{i=1}^n (r_0^2 + a_i^2) \gamma_i
\fft{d\phi_i}{\epsilon_i}\Big]^2\,,\nn
\eea
where $\widetilde U$ and $\widetilde U_\alpha$ are $U$ and $U_\alpha$
given in (\ref{functionsodd}), but with $\hat r$ replaced by $r_0$.
   
   As in the four-dimensional and five-dimensional examples we
discussed in previous sections, we can finally transform from
Poincar\'e coordinates $t$ and $y$ to global coordinates $\tau$ and
$r$, by means of the redefinitions (\ref{poinglob}), together with
appropriate shifts of the azimuthal angles, giving the near-horizon
metric in the form
\bea
ds^2 &=& \fft{\widetilde U}{V} \Big(-(1+r^2) d\tau^2 
  + \fft{dr^2}{1+r^2}\Big)
+\sum_{\alpha=1}^{n-1}
\fft{\widetilde U_\alpha}{X_\alpha} dy_\alpha^2\\
&&+\sum_{\alpha=1}^{n-1}\fft{X_\alpha}{\widetilde U_\alpha}
\Big [\fft{2r_0\prod_{\beta} (r_0^2 + y_\beta^2)}{V (r_0^2 +
y_\alpha^2)} r d\tau +
\sum_{i=1}^n \fft{a_i^2( r_0^2 + a_i^2)\gamma_i}{a_i^2-y_\alpha^2}
\fft{d\phi_i}{\epsilon_i}\Big]^2\nn\\
&&
+\fft{\prod_{k=1}^n a_k^2}{r_0^2 \prod_{\alpha=1}^{n-1} y_\alpha^2}
\Big[\fft{2\prod_\beta (r_0^2 + y_\beta^2)}{r_0 V}
r d\tau + \sum_{i=1}^n (r_0^2 + a_i^2) \gamma_i
\fft{d\phi_i}{\epsilon_i}\Big]^2\,.\nn
\eea

\subsection{$D=2n$ dimensional Kerr-AdS black holes and extremal limit}

  Again we use the same coordinates as those in section 2 of
\cite{chlupo}, but with shifted azimuthal coordinates chosen so that
the metric is simpler, albeit asymptotically rotating.  It is given by
\bea
ds^2&=& \fft{U}{X} d\hat r^2 + \sum_{\alpha=1}^{n-1}
\fft{U_\alpha}{X_\alpha}
dy_\alpha^2 - \fft{X}{U} \Big[d\hat t -
\sum_{i=1}^{n-1} \fft{\gamma_i}{\epsilon_i} d\hat \phi_i\Big]^2
\nn\\
&&+\sum_{\alpha=1}^{n-1} \fft{X_\alpha}{U_\alpha}
\Big[d\hat t - \sum_{i=1}^{n-1} \fft{(\hat r^2+a_i^2)\gamma_i}{
a_i^2 - y_\alpha^2} \fft{d\hat \phi_i}{\epsilon_i}\Big]^2
\,.
\eea
where
\bea
U&=&\prod_{\alpha=1}^{n-1} (\hat r^2 + y_\alpha^2)\,,\qquad
U_\alpha=-(\hat r^2 + y_\alpha^2) {\prod}_{\beta=1}^{'n-1}
(y_\beta^2-y_\alpha^2)\,,\quad
1\le\alpha\le n-1\,,\nn\\
\epsilon_i &=& a_i \Xi_i {\prod}_{k=1}^{'n-1}
(a_i^2-a_k^2)\,,\qquad
\gamma_i=\prod_{\alpha=1}^{n-1} (a_i^2-y_\alpha^2)\,,\quad
1\le i\le n-1\,,\nn\\
X&=&(1 + \hat r^2\ell^{-2})
\prod_{k=1}^{n-1} (\hat r^2 + a_k^2) - 2 M\, \hat r\,,\qquad
\Xi_i =1 - a_i^2 \ell^{-2}\,,\nn\\
X_\alpha&=& -(1-y_\alpha^2\ell^{-2})
\prod_{k=1}^{n-1} (a_k^2 - y_\alpha^2) + 2 L_\alpha\, y_\alpha\,.
\label{functionseven}
\eea
Turning off the NUT parameters, {\it i.e.} $L_\alpha=0$, the solution
describes the Kerr-AdS black hole in even dimensions $D=2n$.  The
coordinate ranges for $y_\alpha$ are somewhat different from the odd
dimensional case.  Without loss of generality, we may order the
rotation parameters such that $a_1\le a_2\le \cdots \le a_{n-1}$, then
we have $-a_1\le y_1\le a_1$, and $a_{\alpha-1} \le y_{\alpha} \le
a_\alpha$ for $\alpha = 2,3,\ldots, (n-1)$.  The $(n-1)$ azimuthal
angles $\hat\phi_i$ each have period $2\pi$.  The asymptotic region is
at $\hat r\rightarrow \infty$, where the metric approaches AdS$_{2n}$
in rotating global coordinates, with $\Omega_i^\infty=-a_i \ell^{-2}$.
The horizon is at $\hat r=r_+$, where $r_+$ is the largest root of the
polynomial function $X$.  The thermodynamic quantities are given by
\bea
T_H&=& 
\fft1{2\pi}\, \Big[r_+(1+r_+^2\ell^{-2})\sum_i \fft{1}{r_+^2 + a_i^2} -
\fft{1 - r_+^2\ell^{-2}}{2r_+}\Big]\,,\nn\\
E &=& \fft{{\cal A}_{D-2}M}{4\pi(\prod_j \Xi_j)} \sum_{i=1}^{n-1}
\fft{1}{\Xi_i}\,,\qquad
S_{BH}=\ft14{\cal A}_{D-2}\prod_{i=1}^{n-1} \fft{r_+^2+a_i^2}{\Xi_i}
\,,\nn\\
J_i&=&\fft{a_i {\cal A}_{D-2}M}{4\pi \Xi_i (\prod_j \Xi_j)}\,,\qquad
\Omega_i=\fft{a_i\Xi_i}{r_+^2 + a_i^2}\,,\label{eventhermo}
\eea
where, as in the previous cases, $\Omega_i$ gives the angular velocities
on the horizon, in the asymptotically-rotating frame.

    As in the case of odd dimensions, the extremal limit is achieved
by choosing parameters such that
\be
X|_{\hat r=r_0}=0\,,\qquad
X'|_{\hat r=r_0}=0\,.
\ee
The horizon is at $\hat r=r_0$, and near the horizon, the function $X$
has the expansion $X=(\hat r-r_0)^2 V + {\cal O}\Big((\hat
r-r_0)^3\Big) $, with $V=\ft12 X''|_{\hat r=r_0}$.

   The near-horizon geometry can be extracted by first
making the coordinate transformation
\be
\hat r = r_0 (1 + \lambda y)\,,\qquad
\hat \phi_i=\phi_i + \alpha_i \hat t\,,\qquad
\alpha_i=\fft{a_i \Xi_i}{r_0^2 + a_i^2}\,,
\label{alphaidef2}
\ee
and the time rescaling
\be
\hat t =\beta t\,,\qquad
\beta=\fft{\prod_i^{n-1} (r_0^2 + a_i^2)}{\lambda r_0 V}\,.
\label{betadef2}
\ee
Finally, taking the limit $\lambda\rightarrow 0$ we obtain the
near-horizon geometry
\bea
ds^2 &=& \fft{\widetilde U}{V} \Big(-y^2 dt^2 + \fft{dy^2}{y^2}\Big)
+\sum_{\alpha=1}^{n-1}
\fft{\widetilde U_\alpha}{X_\alpha} dy_\alpha^2\\
&&+\sum_{\alpha=1}^{n-1}\fft{X_\alpha}{\widetilde U_\alpha}
\Big [\fft{2r_0\prod_{\beta} (r_0^2 + y_\beta^2)}{V(r_0^2+
y_\alpha^2)} y dt +
\sum_{i=1}^{n-1} \fft{( r_0^2 + a_i^2)\gamma_i}{a_i^2-y_\alpha^2}
\fft{d\phi_i}{\epsilon_i}\Big]^2\,,
\eea
where $\widetilde U$ and $\widetilde U_\alpha$ are $U$ and $U_\alpha$ 
as given in
(\ref{functionseven}), but with $\hat r$ replaced by $r_0$.

   In terms of global coordinates on the AdS$_2$, the near-horizon
metric becomes
\bea
ds^2 &=& \fft{\widetilde U}{V} \Big(-(1+r^2) d\tau^2 + 
\fft{dr^2}{1+r^2}\Big)
+\sum_{\alpha=1}^{n-1}
\fft{\widetilde U_\alpha}{X_\alpha} dy_\alpha^2\\
&&+\sum_{\alpha=1}^{n-1}\fft{X_\alpha}{\widetilde U_\alpha}
\Big [\fft{2r_0\prod_{\beta} (r_0^2 + y_\beta^2)}{V(r_0^2 +
y_\alpha^2)} r d\tau +
\sum_{i=1}^{n-1} \fft{( r_0^2 + a_i^2)\gamma_i}{a_i^2-y_\alpha^2}
\fft{d\phi_i}{\epsilon_i}\Big]^2\,.
\eea
(Results for the near-horizon geometries for Myers-Perry black holes 
(\ie with $\ell^{-1}=0$) appear in \cite{kun3}.)

\subsection{Frolov-Thorne vacuum temperature}

There are $[(D+1)/2]$ Frolov-Thorne temperatures associated with the quantum
field theory of the extremal $D$ dimensional metric.
We shall denote these by $T_0$ and
$T_i$, with $1\le i\le n$ or $1\le i\le n-1$ when $D=2n+1$ or $D=2n$
respectively.  Thus we consider
\be
e^{-{\rm i} \omega \hat t + {\rm i} \sum_i m_i \hat \phi_i} =
e^{-{\rm i} n_0 t + {\rm i} \sum_i n_i \phi_i}\,,
\ee
where $(\hat t, \hat \phi_i)$ and $(t, \phi_i)$ are related by
(\ref{alphaidef}) and (\ref{betadef}), for both
even or odd dimensions.  This implies that
\be
n_i=m_i\,,\qquad \omega=\fft{n_0}{\beta} + \sum_i \alpha_i n_i\,.
\ee
Now consider
\be
\exp\Big(-\fft{\omega -\sum_i m_i \Omega_i}{T_H}\Big) =
\exp\Big(-\fft{n_0}{T_0} - \sum_i\fft{n_i}{T_i}\Big)\,.
\ee
Thus we have
\be
T_0=\beta\, T_H\,,\qquad
T_i=\fft{T_H}{\alpha_i-\Omega_i}\,.
\ee
Note that the forms for $\alpha_i$ and $\Omega_i$ are the same for
even and odd dimensions.  In the extremal limit, it is clear that
$T_0=0$.  For the $T_i$, both the numerator and denominator vanish,
leaving a finite and non-vanishing ratio, and hence we find
\be
T_i=\fft{(r_0^2 + a_i^2)^2}{2r_0 a_i \Xi_i}
\Big(\fft{\del T_H}{\del r_+}\Big)\Big|_{r_+=r_0}=
\fft{V r_0^{2\epsilon} (r_0^2 + a_i^2)}{4\pi a_i
\Xi_i r_0 \prod_{j\ne i} (r_0^2+a_j^2)}\,,
\label{tiresult}
\ee
where $\epsilon=0$ and 1 for even and odd dimensions respectively.

\subsection{Central charge and microscopic entropy}

   We may again consider a class of diffeomorphisms analogous to
(\ref{d4diffeo}), which give rise to an asymptotic symmetry algebra of
transformations obeying boundary conditions of the type discussed in
\cite{guhasost}.  In the general cases of rotating AdS black holes in
$D=2n+1$ or $D=2n$ dimensions we shall have $n$ or $(n-1)$ commuting
copies of the Virasoro algebra respectively, generated by the
diffeomorphisms
\be
\zeta_\n^i = - e^{-\im\,n\phi_i}\, \fft{\del}{\del\phi_i} -
   \im\, n\, r\, e^{-\im\, n \phi_i}\, \fft{\del}{\del r}\,.
\ee
The evaluation of the central terms $c_{\phi_i}$ in the Virasoro
algebras at the level of the Dirac brackets of the charges
(\ref{Qdefs}), involving the calculation of the surface integral in
(\ref{centralterm}) using the extension of (\ref{kdef}) to higher
dimensions, is rather complicated and difficult to perform in a
general higher dimension.

   The test of the validity of the Kerr/CFT correspondence in the
higher dimensions would amount to verifying that the central charges
$c_{\phi_i}$ are such that
\be 
S_{BH}= \ft13 \pi^2\, c_{\phi_i}\, T_i\qquad \hbox{for each } i\,,
\label{hdcheck}
\ee
where $S_{BH}$ is the Bekenstein-Hawking entropy of the extremal
rotating AdS black hole, given by (\ref{oddthermo}) or
(\ref{eventhermo}), with $r_+=r_0$ subject to $X(r_0)=0$ and
$X'(r_0)=0$, in odd or even dimensions respectively, and with $T_i$
being the Frolov-Thorne vacuum temperature (\ref{tiresult}) for the
$i$'th Virasoro algebra.

   We have calculated the central charges $c_{\phi_i}$ explicitly in
the cases of $D=6$ and $D=7$ dimensions, and verified that the
relations (\ref{hdcheck}) are indeed satisfied.  Thus in total we have
explicit confirmation of the equality of the microscopic entropy
calculations and the Bekenstein-Hawking entropy for extremal rotating
black holes in dimensions 4, 5, 6 and 7.

\section{Conclusions}

    In this paper, we have extended the recent results in
\cite{guhasost} on the Kerr/CFT correspondence for four-dimensional
Kerr black holes, by including a cosmological constant and also by
considering the equivalent correspondence in all higher dimensions.
The key observation in \cite{guhasost} was that the asymptotic algebra
of a class of diffeomorphisms of the near-horizon geometry of the
extremal Kerr black hole, subject to certain boundary conditions, contains 
a Virasoro algebra related to reparameterisations of the azimuthal
coordinate $\phi$ in the Kerr metric.  This algebra could be
associated with a chiral CFT, and, realised at the level of Dirac
brackets of diffeomorphism charges, it has a central term related to
the angular momentum of the black hole.  It was shown in
\cite{guhasost} that the microscopic entropy of the CFT, calculated
using the Cardy formula, coincides with the Bekenstein-Hawking entropy
of the extremal Kerr black hole.

   In dimensions higher than 4, the generalisations of the Kerr and
Kerr-AdS metrics have multiple independent angular momenta, associated
with rotations in mutually-orthogonal spatial 2-planes
\cite{myper,hawhuntay,glpp1,glpp2}.  This leads to an asymptotic
symmetry that includes multiple mutually commuting copies of the
Virasoro algebra.  The central terms in these Virasoro algebras are
all different (when the rotation parameters are unequal), but
nevertheless, when we evaluate the microscopic entropy for each of the
associated chiral CFTs using the Cardy formula, we find that it agrees
precisely with the Bekenstein-Hawking entropy of the extremal Kerr-AdS
black hole.

   It is intriguing to note that in the case of rotating black holes
with a (negative) cosmological constant, there are two ostensibly very
different types of duality that can be considered.  For asymptotically AdS
black holes, one expects that the Bekenstein-Hawking entropy can be
calculated in the boundary field theory {\it via} the AdS/CFT correspondence
\cite{mald,guklpo,wit}.  In particular, in five dimensions one can
consider the duality of the Kerr-AdS$_5$ black hole to a
four-dimensional rotating boundary field theory \cite{hawhuntay}.  (In
the case of supersymmetric charged rotating black holes in five
dimensions \cite{gutrea,cclp}, a boundary free-fermion approximation was
used to obtain the entropy up to a numerical factor of order unity
\cite{kmmr}.) On the other hand, as we have discussed in this paper,
there is a Kerr/CFT correspondence in the near-horizon region, in the
case of extremal black holes.  It would be interesting to see whether
this multiplicity of dualities persists in a more general context.

\section*{Acknowledgements}

  Research  supported in part by 
DOE grant DE-FG03-95ER40917.  We are grateful to David Chow for
useful discussions.

\end{document}